\documentclass[12pt]{article}
\usepackage{hyperref,epsfig,amssymb,amsmath,graphicx,color}

\begin{document}

\baselineskip=.22in
\renewcommand{\baselinestretch}{1.2}
\renewcommand{\theequation}{\thesection.\arabic{equation}}
\newcommand{\klmt}{\mbox{K\hspace{-7.6pt}KLM\hspace{-9.35pt}MT}\ }
\begin{flushright}
{\tt  }
\end{flushright}

\vspace{1mm}

\begin{center}
{{\Large \bf Solutions in
IR modified Ho${\check {\bf r}}$ava-Lifshitz Gravity}\\[20mm]
Taekyung Kim\\[2mm]
{\it Department of Physics, BK21 Physics Research Division,
and Institute of Basic Science,}\\
{\it Sungkyunkwan University, Suwon 440-746, Korea}\\
{\tt pojawd@skku.edu}\\[7mm]

Chong Oh Lee\\[2mm]
{\it Department of Physics and Astronomy,,}\\
{\it University of Waterloo, Waterloo, Ontario, N2L 3G1, Canada}\\
{\tt cohlee@sciborg.uwaterloo.ca}}
\end{center}

\vspace{8mm}

\begin{abstract}
In order to allow the asymptotically flat, we consider Ho\v{r}ava-Lifshitz gravity theory
with a soft violation of the detailed balance condition
and obtain various solutions. In particular, we find that
such theory coupled to a global monopole leads to a solution representing a space
with deficit solid angle, which is well matched with genuine feature of GR.
\end{abstract}


\newpage

\setcounter{equation}{0}
\section{Introduction}\label{sec1}
The construction of the ultra-violet(UV) complete theory of gravity
has been an intriguing subject of discussions for theoretical
physics of the past fifty years. The discussion has been recently
concentrated on the UV complete theory in space and time with an
anisotropic scaling in a Lifshitz fixed
point~\cite{Lifshitz,Horava:2008jf,Horava:2008ih,Horava:2009uw,Horava:2009if}.
In particular, this theory is very attractive since pertubative
renormalizability is realized as well as Lorentz symmetry is
recovered in low energy regime in spite of being broken the Lorentz
symmetry in high energy.

Ho${\check {\bf r}}$ava-Lifshitz gravity (HL) has been studied in
various directions, which are categorized into two. One is investigating
and developing the properties of the HL theory itself~
\cite{Visser:2009fg}--\cite{Blas:2009ck}. The other is
applying this theory to cosmological framework including the black hole
solutions~\cite{Lu:2009em}--\cite{Greenwald:2009kp}
and their thermodynamic prosperities~\cite{Myung:2009dc}--\cite{Majhi:2009xh}.

The metric in the (3+1)-dimensional ADM decomposition can be written
as
\begin{align}
ds^{2}=-N^2dt^2+g_{ij}\left(dx^i+N^idt\right)\left(dx^j+N^jdt\right),
\label{met1}
\end{align}
where $N(t,x^{i})$ denotes the lapse function, $g_{ij}(t,x^{i})$ is
the spatial metric, and $N_{i}(t,x^{i})$ is the shift function.
Then, the Einstein-Hilbert action can be expressed as
\begin{align}\label{EHa}
S_{\rm EH}=\frac{1}{16\pi G}\int d^{4}x\sqrt{g}N(K_{ij}K^{ij}-K^2+R-2\Lambda),
\end{align}
where $G$ is Newton's constant and the extrinsic curvature for a spacelike
hypersurface with a fixed time is
\begin{align}
K_{ij}\equiv
\frac{1}{2N}\left({\dot{g_{ij}}}-\nabla_iN_j-\nabla_jN_i\right).
\end{align}
Here, a dot denotes a derivative with respect to $t$ and covariant derivatives
defined with respect to the spatial metric $g_{ij}$.

The IR-modified HL action with asymptotically flat limit is given by
~\cite{Horava:2009uw,Nastase:2009nk,Kehagias:2009is}
\begin{align}\label{ac1}
S_{{\rm HL}} =&\int dt\, d^{3}x\,\sqrt{g} N({\cal L}_{{\rm IR}}+{\cal
L}_{{\rm UV}}),\\
{\cal L}_{{\rm IR}}=&{2\over \kappa^2}(K_{ij}K^{ij}-\lambda K^2)
{+\frac{\kappa^2\mu^2 }{8(1-3\lambda)}
\left[\left(\Lambda-\omega \right)R-3\Lambda^2\right]},
\label{LIR}\\
{\cal L}_{{\rm UV}}=&-{\kappa^2\over 2\nu^4}\left(C_{ij}-\frac{\mu
\nu^{2}}{2}R_{ij}\right) \left(C^{ij}-\frac{\mu \nu^{2}}{2}R^{ij}\right)
+\frac{\kappa^2\mu^2(1-4\lambda)}{32(1-3\lambda)}R^2, \label{LUV}
\end{align}
where $R$ and $R_{ij}$ are three-dimensional scalar curvature and  Ricci tensor, and
the Cotton tensor is given by
\begin{align}
C^{ij}=
\frac{\epsilon^{ikl}}{\sqrt{g}}\nabla_k\left({R^j}_l-\frac{1}{4}R\delta^{j}_{\;l}\right).
\label{Co1}
\end{align}
The action has parameters, $\kappa,\lambda,\nu,\mu,\Lambda,$ and $\omega$.
In the limit of vanishing cosmological constant $\Lambda\rightarrow 0$, one compares the IR-modified action
(\ref{ac1}) with the (3+1)-dimensional Einstein-Hilbert action (\ref{EHa}) and
reads the parameter $\lambda$, the speed of light $c$, Newton's constant $G$ as
\begin{align}
\lambda=1,\quad
c^2=\frac{\kappa^4\mu^2\omega}{32},\quad
G= \frac{\kappa^2}{32\pi c}.
\label{con1}
\end{align}

Recently, HL gravity coupled to electrostatic field of a point charge is considered and an exact
solution is found, describing a space with either a surplus or
deficit solid angle is found~\cite{Kim:2009dq}. The surplus angle due to an ordinary matter
with positive energy density in~\cite{Kim:2009dq} is not well matched with known result
of GR in which it can usually be materialized by the source of negative mass or
energy. However, from cosmological point of view, one finds
the detailed balance condition leads to obstacles~\cite{Nastase:2009nk,Calcagni:2009qw}.
Furthermore by introducing
a soft violation of the detailed balance condition, they show that their results
are consistent with them of GR~\cite{Kehagias:2009is}.
Thus one intriguing question is whether IR-modified HL theory coupled to matter field
reproduces them of GR.

In this paper, we address this question. We consider IR-modified HL in presence of
the global monopole, and find a spherically symmetric solution describing a space with
deficit solid angle.

The paper is organized as follows. In section 2, vacuum solutions are discussed under spherical
symmetry. In section 3, we obtain the deficit solid angle due to the solution of IR modified HL
gravity with the global monopole. Finally, we give a conclusion.

\setcounter{equation}{0}
\section{Vacuum Solutions under Spherical Symmetry}\label{sec2}
Let us investigate a spherically symmetric solution with the static metric ansatz
\begin{equation}
ds^2=-{\cal F}(r)e^{2\rho(r)}dt^2+\frac{dr^2}{{\cal F}(r)}+r^2(d\theta^2+\sin^2\theta
d\varphi^2). \label{rmet}
\end{equation}
Since all the components of Cotton tensor vanish under this metric,
the action (\ref{ac1}) reduces to
\begin{align}
S_{{\rm HL}} =& 4\pi \int_{-\infty}^{\infty}dt\int_{0}^{\infty} dr
r^2 \,e^{\rho} \Bigg\{
-\frac{\kappa^2\mu^2}{8}\left[\left(\frac{{\cal F}'}{r}\right)^2
+\frac{2}{r^4}\left(1-{\cal F}-\frac{r{\cal F}'}{2}\right)^2\right]\nonumber\\
&\hspace{30mm}+\frac{\kappa^2\mu^2}{8(1-3\lambda)}
\left[\frac{1-4\lambda}{r^4}(1-{\cal F}-r{\cal F}')^2+{\frac{
2(\Lambda-\omega)}{r^2}(1-{\cal F}-r{\cal F}')}
 -3\Lambda^2\right]\Bigg\}
\nonumber\\
=&  \frac{\pi\kappa^2\mu^2}{2(3\lambda-1)} \int dt\int dr \,
e^{\rho}\times
\nonumber\\
&\Bigg\{ (1-3\lambda) \Bigg[{\tilde {\cal F}}'^2 +2\Big(\frac{\tilde {\cal F}}{r}
+\frac{{\tilde {\cal F}}'}{2}\Big)^2 \Bigg] -(1-4\lambda)\Big(\frac{{\tilde
{\cal F}}}{r}+{\tilde {\cal F}}'\Big)^2 {+2(\Lambda-\omega) r\Big(\frac{{\tilde
{\cal F}}}{r}+{\tilde {\cal F}}'\Big)}+3\Lambda^2r^2 \Bigg\}, \label{rac}
\end{align}
where ${\tilde {\cal F}}={\cal F}-1$.
Then, the equations of motion are obtained as
\begin{align}
&~\Bigg[(\lambda-1){\tilde {\cal F}}'-\frac{2\lambda}{r}{\tilde {\cal F}}
{-2(\Lambda-\omega) r}\Bigg]\rho'+ (\lambda-1)\tilde {\cal F}'' -
\frac{2(\lambda-1)}{r^{2}}\tilde {\cal F}
=0,
\label{Beq}\\
&~ (1-3\lambda) \Bigg[ {\tilde {\cal F}}'^2+2\Big( \frac{\tilde {\cal F}}{r}
+\frac{\tilde {\cal F}'}{2} \Big)^2 \Bigg] -(1-4\lambda)\Big(\frac{\tilde
{\cal F}}{r}+{\tilde {\cal F}}'\Big)^2 {+2(\Lambda-\omega) r}
\Big(\frac{\tilde {\cal F}}{r}+{\tilde
{\cal F}}'\Big)+3\Lambda^2 r^{2}=0. \label{deleq}
\end{align}

We start by giving a brief discussion of the asymptotic behaviors of
the solutions to Eqs.~(\ref{Beq}) and (\ref{deleq}).
In the low energy regime, taking the $\lambda=1$ and
neglecting the quadratic terms in the metric functions,
the equations~(\ref{Beq})--(\ref{deleq}) reduce to the Einstein equations,
which reproduce Schwarzchild solution
in the limit $\Lambda \rightarrow 0$ as we expect
\begin{align}
r\frac{d\rho}{dr}=0, \quad\longrightarrow&\quad
\rho(r)=\rho_0=0,
\label{Beq2}\\
\frac{d}{dr}\left(r{\cal F}\right)=1,
\quad~\longrightarrow&\quad
{\cal F}(r)=1-\frac{M}{r}, \label{deleq2}
\end{align}
where $M$ is an integration constant.

For sufficiently large $r$ at asymptotic region, it is assumed that
the divergence of ${\cal F}(r)$ arises as a power behavior.
A straightforward calculation with Eq. (\ref{deleq}) leads to
\begin{align}
{\cal F}(r)\approx \left\{
\begin{array}{cll}
({\rm I}) &(\omega-\Lambda)r^2-\sqrt{\omega(\omega-2\Lambda)}\,r^2
& \mbox{for~ arbitrary~~}\lambda \\
({\rm II}) & {\displaystyle {\cal F}_{\rm IR}r^{p}} &
\mbox{for~~}\displaystyle{\lambda>1}
\end{array}
\right. , \label{ccso}
\end{align}
where the coefficient ${\cal F}_{\rm IR}$ is an undetermined constant and
\begin{align}
p=\frac{2\lambda +\sqrt{2(3\lambda -1)}}{\lambda -1}\ .
\label{p}
\end{align}
It is shown that the behavior of the long distance in (I) without a
cosmological constant agrees with that of the leading IR behavior
in (\ref{deleq}). The long distance behavior in (II) seems to imply a new possible
solution which comes from higher derivative terms.

For sufficiently small $r$ at the UV regime, assuming the divergence of $B(r)$ follows
as power behavior
\begin{align}
{\cal F}(r)\sim\frac{\beta}{r^{l}},\qquad (\beta={\rm constant},~l>0),
\end{align}
the leading term in Eq.~(\ref{deleq}) is proportional to $1/r^{2l+2}$.
The contribution to the correction term
due to the soft violation of the detailed balance condition in Eq.~(\ref{deleq})
can be neglected since such contribution is proportional to $1/r^{l}$. Thus,
the leading UV behavior in IR modified HL theory is exactly the same as that in HL theory.
The allowed powers for various $\lambda$ are given as

\begin{align}
{\cal F}(r)\approx \left\{
\begin{array}{cll}
({\rm A}) & 1 & \mbox{for~ arbitrary~~}\lambda \\
({\rm B}) & b &
\mbox{for}~~\displaystyle{\lambda=\frac{1}{2}} \\
({\rm C}) & {\cal F}_{\rm UV+} r^{p}~~\mbox{or}~~{\cal F}_{\rm UV-}r^{q} &
\mbox{for}~~\displaystyle{\frac{1}{3}\le
\lambda<\frac{1}{2}} \\
({\rm D}) & {\cal F}_{\rm UV+} r^{p} &
\mbox{for}~~\displaystyle{\frac{1}{2}<\lambda<1}
\end{array}
\right. , \label{ccsi}
\end{align}
where $b$ denotes an integration constant,
$B_{\rm UV\pm}$ are undetermined constants, $p$ is given in \eqref{ccso},
and $q$ is
\begin{align}
q=\frac{2\lambda -\sqrt{2(3\lambda -1)}}{\lambda -1}\ .
\label{q}
\end{align}

We show that we find new exact vacuum solutions and discuss
how they connect two asymptotes with various value of $\lambda$.
For arbitrary $\lambda$, a solution to the equations~(\ref{Beq})--(\ref{deleq})
obtained as
\begin{align}
{\cal F}=1+(\omega-\Lambda)r^2-\sqrt{\omega(\omega-2\Lambda)}\,r^2, \qquad \rho=\rho_0=0,
\label{vso3}
\end{align}
which connects (I) and (A).
For $\lambda=1/3$, another static exact solution is
\begin{align}
{\cal F}=1+(\omega-\Lambda)r^2-\sqrt{\omega(\omega-2\Lambda)}\,r^2-\frac{M}{r},
\qquad \rho=\rho_0=0,
\label{vso2}
\end{align}
which reproduces AdS Schwarzschild black hole solution with twice cosmological constant
for $\omega=0$. This result in IR modified HL theory agrees with that
in HL~\cite{Lu:2009em,Kim:2009dq}.
For $\lambda=1$, the known exact solution is obtained by~\cite{Park:2009zra}
\begin{align}
{\cal F}=1+(\omega-\Lambda)r^2-\sqrt{\omega(\omega-2\Lambda)r^4+c\,r}, \qquad \rho=\rho_0=0,
\label{vso1}
\end{align}
where $c$ is an integration constant. This solution also connects (I) and (A).

In contrast to the exact vacuum solutions in HL theory~\cite{Lu:2009em,Kim:2009dq},
it is not clear how they have connection between (\ref{ccso}) and (\ref{ccsi})
since it seems that there are not other exact solutions in IR modified HL theory except
previous exact solutions (\ref{vso3})--(\ref{vso1}), i.e., there do not exist
exact solutions with covering all range of $\lambda$ for $\lambda\geq1/3$.
It presumably implies that all the vacuum solutions
in IR modified HL theory do not always follows as power behavior.

Horizons and singularities in HL gravity
have been discussed in the previous work~\cite{Kim:2009dq}.
However, we do not deal with them since HL theory does not have full diffeomorphism
invariance and both of the previous concepts are not easy to discern~\cite{Kiritsis:2009rx}.

\section{Global Monopole Solution}\label{sec2}
In the presence of matter field, it is described by action
\begin{align}\label{acma}
S_{\rm m}&=\int dtd^3x \sqrt{g}N~{\cal L}_{{\rm m}}(N,N_i,g_{ij})\\
&=4\pi \int_{-\infty}^{\infty}dt\int_{0}^{\infty} dr r^2
e^{\rho} {\cal L}_{\rm m}({\cal F},\rho).
\end{align}
Then, the equations of motion are given by
\begin{align}
&~\Bigg[(\lambda-1){\tilde {\cal F}}'-\frac{2\lambda}{r}{\tilde {\cal F}}
{-2(\Lambda-\omega) r}\Bigg]\rho'+ (\lambda-1)\tilde {\cal F}'' -
\frac{2(\lambda-1)}{r^{2}}\tilde {\cal F}
=\frac{8(1-3\lambda)r^2}{\kappa^2\mu^2}\frac{\partial {\cal L}_{\rm
M}}{\partial {\cal F}},
\label{Beq2}\\
&~ (1-3\lambda) \Bigg[ {\tilde {\cal F}}'^2+2\Big( \frac{\tilde {\cal F}}{r}
+\frac{\tilde {\cal F}'}{2} \Big)^2 \Bigg] -(1-4\lambda)\Big(\frac{\tilde
{\cal F}}{r}+{\tilde {\cal F}}'\Big)^2 {+2(\Lambda-\omega) r}
\Big(\frac{\tilde {\cal F}}{r}+{\tilde
{\cal F}}'\Big)+3\Lambda^2 r^{2}
\nonumber\\
&\hspace{85mm}=\frac{8(1-3\lambda)r^{2}}{\kappa^2\mu^2}\left({\cal
L}_{\rm m} +\frac{\partial {\cal L}_{\rm m}}{\partial
\rho}\right). \label{deleq2}
\end{align}

When we consider a global monopole of O(3) linear sigma model and magnetic monopole of
U(1) gauge theory in the HL type field theory, the long distance behavior of the
Lagrangian density in IR regime must be proportional to $1/r^{n}$ irrespective of the
value of $z$ (see Ref.~\cite{Kim:2009dq} for more details)
\begin{align}
\frac{\partial {\cal L}_{\rm m}}{\partial {\cal F}}\approx 0,\qquad {\cal
L}_{\rm m}+\frac{\partial {\cal L}_{\rm m}}{\partial \rho} \approx
-\frac{\gamma}{r^{n}}, \qquad (n=0,1,2,...), \label{M4}
\end{align}
where a constant $\gamma$ is determined by the explicit Lagrangian form
and the monopole configurations of interest.
Positive $\gamma$ can be read off from the energy momentum tensor
of matter fields
and $n$ must be a positive integer in order to get a finite energy.
A straightforward calculation with Eqs. (\ref{Beq2}) and (\ref{deleq2}) leads to
\begin{align}
{\cal F}&=1+\left[(\omega-\Lambda)\pm\sqrt{\omega(\omega-2\Lambda)}\right]r^2
+\frac{8(n-3)\gamma}{n^2\kappa^2\mu^2\sqrt{\omega(\omega-2\Lambda)}}r^{2-n},\\
\rho&=(2n-3)\ln (r/r_0)+\left(\frac{3}{n}-2\right)\ln
\left[\frac{8\gamma(n-3)^2}{\kappa^2\mu^2}-\omega(\omega-2\Lambda)n^3r^n\right],
\label{mso2}
\end{align}
for $n\neq3$ and $\lambda={(n^2-4n+6)}/{n^2}$.
In particular, in the case of $n=3$, there exists solution only by taking $\lambda=1/3$.
Then, matter contributions vanish in \eqref{Beq2} and \eqref{deleq2} when $\lambda=1/3$.
Therefore, such solution exactly goes back to the vacuum solution \eqref{vso2}.
One also finds special solution for $\lambda=1$,
\begin{align}
{\cal F}&=1+(\omega-\Lambda)r^2\pm\sqrt{\omega(\omega-2\Lambda)r^4+f\,r
+\frac{16\gamma}{(3-n)\kappa^2\mu^2}r^{4-n}}, \qquad \rho=\rho_0=0, \qquad (n\neq3)\\
{\cal F}&=1+(\omega-\Lambda)r^2\pm\sqrt{\omega(\omega-2\Lambda)r^4+f\,r
+\frac{16\gamma}{\kappa^2\mu^2}r\ln r}, \qquad \rho=\rho_0=0, \qquad (n=3)
\label{mso1}
\end{align}
with an integration constant $f$.

Let us study the details of the global monopole solution.
The O(3) sigma model action is presumably taken as
\begin{align}
S_{{\rm O(3)}}=\int d^{4}x\sqrt{-g_{4}}\left(-\frac{g^{00}}{2}
\partial_{0}\psi^{a}\partial_{0}\psi^{a} -V\right),
\label{Oir}
\end{align}
where, $\psi^{a}$~$(a=1,2,3)$ denote a scalar fields and $g^{00}=1/N^{2}$.
For simplicity, we assume an ordinary quadratic spatial derivatives and
of a quartic order self-interactions,
\begin{align}
V(\psi^{a},\partial_{i}\psi^{a},...)
=-\frac{g^{ij}}{2}\partial_{i}\psi^{a}\partial_{j}\psi^{a}
-\frac{\lambda_{{\rm m}}}{4}(\psi^{2}-v^{2})^{2},\qquad
\psi^{2}\equiv \psi^{a}\psi^{a},
\label{VIR}
\end{align}
For anisotropic scaling $z=1$ $(n=2)$, the IR action (\ref{Oir}) is
\begin{align}
S_{\rm O(3)}=4\pi\int_{-\infty}^{\infty}dt\int_{0}^{\infty}dr r^2
e^{\rho}\left[-\frac{{\cal F}}{2}\psi'^{\,2}-\frac{\psi^2}{r^2}
-\frac{\lambda_{{\rm m}}}{4}(\psi^2-v^2)^2\right] ,
\end{align}
and, under a hedgehog ansatz
\begin{align}
\psi^{a}={\hat r}^{a}\psi(r)=(\sin\theta\cos\varphi,\,
\sin\theta\sin\varphi,\, \cos\theta) \psi(r),
\label{mnan}
\end{align}
it leads to
\begin{align}
\frac{\partial {\cal L}_{\rm m}}{\partial {\cal F}}=&
-\frac{1}{2}\psi'^{\,2},
\label{mat1}\\
{\cal L}_{\rm m}+\frac{\partial {\cal L}_{\rm m}}{\partial \rho}=&
-\frac{{\cal F}}{2}\psi'^{\,2}-\frac{\psi^2}{r^2}-\frac{\lambda_{{\rm
m}}}{4}(\psi^2-v^2)^2. \label{mat2}
\end{align}
Two boundary conditions of the above equations are imposed by requiring single-valuedness of the
field at the monopole position and finite energy at spacial infinity
\begin{align}
\psi(0)=0,\qquad \psi(\infty)=v. \label{scbc}
\end{align}
From the boundary conditions, one can take the following configuration
\begin{align}
\psi(r)=
\begin{cases}
0,&~ \mbox{for}~~ r\le \displaystyle \frac{1}{v\sqrt{\lambda_{{\rm m}}}},\\
v,&~ \mbox{for}~~ r>\displaystyle \frac{1}{v\sqrt{\lambda_{{\rm m}}}},
\end{cases}
\end{align}
which means the scalar field $\psi(r)$ has vacuum expectation value zero in the region
inside the monopole core and $v$ outside, respectively.
Therefore, the field equations (\ref{mat1})--(\ref{mat2}) near the vacuum reduce to
\begin{align}
\frac{\partial {\cal L}_{\rm m}}{\partial {\cal F}}\approx  0, \qquad {\cal
L}_{\rm m}+\frac{\partial {\cal L}_{\rm m}}{\partial \rho}\approx
-\frac{v^{2}}{r^{2}}.
\label{mat4}
\end{align}
In particular, $\gamma$ in (\ref{M4}) is given as $v^2$ for $n=2$.
Then, the metric function ${\cal F}(r)$ is obtained by
\begin{align}
{\cal F}&=1+\left[(\omega-\Lambda)\pm\sqrt{\omega(\omega-2\Lambda)}\right]r^2
-\frac{2v^2}{\kappa^2\mu^2\sqrt{\omega(\omega-2\Lambda)}},\\
\rho&=\ln (r/r_0)-\frac{1}{2}\ln\left[\frac{v^2}{\kappa^2\mu^2}-\omega(\omega-2\Lambda)r^2\right],
\label{mso3}
\end{align}
which leads to
\begin{align}
ds^{2}=&-\frac{1+\left[(w-\Lambda)\pm\sqrt{w(w-2\Lambda)}\right]r^2}
{\frac{v^2}{\kappa^2\mu^2}+\left[-w(w-2\Lambda)+\frac{2v^2}{\kappa^2\mu^2}\sqrt{w(w-2\Lambda)}\right]r^2}dt^2
+\frac{dr^2}{1+\left[(w-\Lambda)\pm\sqrt{w(w-2\Lambda)}\right]r^2}\nonumber\\
&+r^{2}\left(1- \frac{2v^2}{\kappa^2\mu^2\sqrt{w(w-2\Lambda)}}\right) (d\theta^2+\sin^2\theta d\varphi^2),\label{met2}
\end{align}
after rescaling the coordinates,
\begin{align}
dt\rightarrow \left(1-\frac{2v^2}{\kappa^2\mu^2\sqrt{w(w-2\Lambda)}}\right)^{-1}r_0 dt, \qquad dr\rightarrow
\sqrt{1-\frac{2v^2}{\kappa^2\mu^2\sqrt{w(w-2\Lambda)}}}\, dr.
\end{align}
The metric \eqref{met2} describes a space with a deficit solid angle
~\cite{Barriola:1989hx,Kim:1996pa,Kim:2009dq},
\begin{align}
4\pi\Delta=\frac{8\pi v^2}{\kappa^2\mu^2\sqrt{w(w-2\Lambda)}},\qquad \mbox{for}~~ 0<\frac{2v^2}{\kappa^2\mu^2\sqrt{w(w-2\Lambda)}}<1.
\end{align}
In \eqref{met2} a black hole horizon is formed at
\begin{align}
r_{\rm H}=\frac{\frac{2v^2}{\kappa^2\mu^2\sqrt{w(w-2\Lambda)}}-1}{\sqrt{(w-\Lambda)\pm\sqrt{w(w-2\Lambda)}}}
\qquad \mbox{for}~~ \frac{2v^2}{\kappa^2\mu^2\sqrt{w(w-2\Lambda)}}\geq1.
\end{align}
These results show two genuine features of GR;
there does not exist a surplus but deficit solid angle and a source
which gives rise to deficit angle is not an electric field
but a scalar field.

In this section, we concentrate on investigating a solid angle in low
energy limit. One can also examine other issues such as a potential in the
UV action and energy configurations near the Lifishitz fixed point as in
~\cite{Kim:2009dq}.

\setcounter{equation}{0}
\section{Conclusion}\label{sec5}

We introduce HL gravity theory with a soft violation of the detailed
balance condition with/without matter fields of power-law behaviors as $1/r^{n}$
and find various solutions.
The IR-modified HL theory coupled to matter field
for $n = 2$ is of particular interest since such theory has only the deficit solid angle
and source giving rise to deficit angle is the scalar field, which agree
with well known results of GR. It seems to imply the detailed
balance condition should be violated if one applies HL theory as cosmological fame works
and wants to obtain the realistic cosmological results in our universe.

\section*{Acknowledgments}
We would like to thank R.B. Mann for useful discussions and Yoonbai
Kim for helpful comments on the manuscript. This work was
supported by the Korea Research Foundation Grant funded by the
Korean Government (KRF-2008-357-C00018).

\end{document}